\documentclass{article}
\usepackage{spconf,amsmath,graphicx,hyperref}
\usepackage{multirow}
\usepackage{adjustbox}
\usepackage{color}
\usepackage{tabularx} % 导言区加上
\usepackage{enumitem} % 在导言区加入
\usepackage{makecell} % 导言区
\usepackage{setspace}
\usepackage[sorting = none, backend = bibtex, style=numeric-comp]{biblatex}
\AtBeginBibliography{\small}
\title{Physics-Informed Video Diffusion for Shallow Water Equations}
\name{Yang Bai$^{1,2}$ \qquad George Eskandar$^{3*}$\thanks{* : Now at Tavus.} \qquad Ziyuan Liu$^{3{\dagger}}$\thanks{${\dagger}$ : Corresponding author.} \qquad Gitta Kutyniok$^{1,2,4,5}$}
\address{$^{1}$ Department of Mathematics, Ludwig-Maximilians-Universität München (LMU Munich), Germany\\
$^{2}$ Munich Center for Machine Learning (MCML), Munich, Germany\\
$^{3}$ Huawei Heisenberg Research Center (Munich), Germany\\
$^{4}$ Institute of Robotics and Mechatronics, DLR-German Aerospace Center, Germany\\
$^{5}$ Department of Physics and Technology, University of Tromsø, Tromsø, Norway}

\begin{document}
%\ninept
%
\maketitle
\begin{abstract}
% Physics-based simulations are widely used in computer graphics to generate realistic animations, but they are computationally expensive. 
Traditional fluid dynamics simulation pipelines combine numerical solvers with rendering, producing highly realistic results but at considerable computational cost. Diffusion-based generative video models offer a faster alternative, yet often ignore physical laws and thus fail to capture consistent dynamics. We propose a physics-informed video diffusion framework that jointly generates visual outputs and physical states. Unlike prior two-stage approaches that first simulate the physical variables and then render, we directly integrate physics constraints into the generative process, enabling simultaneous prediction of physical states and realistic videos without a separate rendering step. Built on the two-dimensional shallow water equations with terrain topography, our method produces temporally coherent water flow while maintaining physical plausibility. Experiments show that it outperforms purely data-driven video diffusion baselines in both realism and physical fidelity, while generating videos significantly faster than traditional simulation-plus-rendering pipelines.

\end{abstract}
\vspace{-1em} % 可选，调整上下间距

\section{Introduction}
\label{sec:intro}
In traditional computer graphics (CG) pipelines, fluid visuals are generated through a two-stage process: first, a physics-based simulator computes physical states; second, these states are passed to a rendering module to produce the final images or videos. Classical solvers based on meshes or particles can reproduce complex behaviors such as turbulence, splashing, and soft-body deformation~\cite{stam2023stable, foster1996realistic, thurey2010multiscale, wojtan2010physics, premvzoe2003particle, solenthaler2009predictive}. These simulations produce physically plausible motion. However, the additional step of photorealistic rendering for fine-scale surface details, shading, and lighting substantially increases computational costs. Consequently, producing a single high-resolution sequence can take hours or even days, making such pipelines impractical for interactive or large-scale applications.
 
To meet the demands of real-time graphics applications like game engines, the industry often employs spectrum-based and FFT-based water rendering techniques~\cite{johanson2004real}. Instead of a physically accurate solution to the Navier-Stokes equations~\cite{navierstokes}, these methods use statistical approximations. This trade-off prioritizes efficiency and visual plausibility over strict physical accuracy. In contrast, domains requiring strict physical fidelity, such as scientific research or film visual effects still rely on high-fidelity Navier-Stokes solvers. Implemented in software like Clawpack~\cite{mandli2016clawpack}, OpenFOAM~\cite{jasak2009openfoam}, or Basilisk~\cite{kenneally2020basilisk}, such solvers demand prohibitive computational resources, especially at high-grid resolutions or large scales. A range of acceleration strategies has been explored to address these challenges: for example,~\cite{kochkov2021machine} refine low-grid resolution simulations, while Physics-Informed Neural Networks (PINNs) method~\cite{cai2021physics} accelerate the solution of specific Partial Differential Equations (PDEs) terms. Nevertheless, a fundamental bottleneck persists: the combination of physics-based simulation and photorealistic rendering remains computationally expensive. The initial high cost of solving the physical equations, even when accelerated, is compounded by the subsequent time-intensive rendering process.

Recently, diffusion-based video generation methods~\cite{ho2022video, ho2020denoising} have emerged as a promising alternative for dynamic scene visualization. Unlike classical pipelines, these methods synthesize video sequences directly, bypassing explicit physical simulation and costly rendering. Crucially, their generation speed is remarkably fast, depending solely on the number of sampling steps during inference, regardless of the scene's complexity. By learning spatio-temporal patterns from large-scale datasets, they can produce visually convincing motions and textures even in complex scenarios. However, because the generated videos are not constrained by the physical laws of the real world, they frequently exhibit temporal inconsistencies and behaviors that violate basic physical principles. This limitation becomes especially pronounced for phenomena governed by complex physical equations. In such situations, especially in fluid dynamics, reproducing realistic and stable motion is extremely challenging for diffusion-based models without explicit physical guidance.\\
% \textbf{Contribution:} We propose a physics-informed video generation framework that tightly integrates grid-based numerical solvers with diffusion models. By embedding the shallow water equations (SWEs) and terrain topography directly into the generative pipeline, our method jointly produces video sequences and corresponding physical states. This design bypasses costly rendering while maintaining both high visual quality and physical consistency, resulting outputs are temporally stable, physically interpretable. Empirical results demonstrate that our framework generates more realistic fluid motion than purely data-driven baselines, and its generation speed is dramatically faster than traditional simulation-plus-rendering pipelines. This effectively bridges the gap between simulation accuracy and generative efficiency.
\textbf{Contributions} To address the limitations of purely data-driven video diffusion models, we propose a physics-informed video generation framework that tightly combine grid-based numerical methods with diffusion models. By embedding the initial conditions of the shallow water equations (SWEs) and terrain information into the generative pipeline, our model jointly produces video sequences and corresponding physical states. Specifically, our contributions are:

% 文中 enumerate
\begin{enumerate}[itemsep=0em, topsep=0em, parsep=0em, partopsep=0em, leftmargin=*]
% \item Our framework co-generates video frames and physical states simultaneously, ensuring that the generated videos adhere to the underlying fluid dynamics.
\item We present the first framework that co-generates video frames and physical states, ensuring that the generated videos adhere to the underlying fluid dynamics.
\item We incorporate the SWEs and terrain directly into the diffusion transformer, bypassing costly rendering while maintaining high visual quality, temporal stability, and physical interpretability.
\item Compared to classical simulation-plus-rendering pipelines, our method achieves over an order of magnitude reduction in runtime, with performance largely unaffected by grid resolution. Despite this speedup, it preserves between $67\%$ to $90\%$ of the simulation accuracy, while producing more realistic fluid motion than purely data-driven baselines, effectively bridging the gap between physical fidelity and generative efficiency.
% \item Compared to classical simulation-plus-rendering pipelines, our method achieves dramatically faster inference, with runtime largely unaffected by grid resolution, while producing more realistic fluid motion than purely data-driven baselines, effectively bridging the gap between simulation accuracy and generative efficiency.
\end{enumerate}

\vspace{-1em} % 可选，调整上下间距
\section{Related Work}
\label{sec:relatedwork}

\vspace{-0.5em} % 可选，调整上下间距

%, achieving visual realism does not guarantee true physical understanding. Consequently, these models often produce videos that do not adhere to underlying physical rules. 

% NOT NEEDED IN RELATED WORKS TOO MANY DETAILS For instance,~\cite{cao2024teaching} pretrains a masked autoencoder (MAE) to capture latent physical knowledge, while~\cite{lin2025reasoning} generates pseudo-linguistic prompt features via CLIP alignment to guide video generation. DiffPhy~\cite{chen2024diffphys}, GPT4Motion~\cite{lv2024gpt4motion}. Similarly PhysGen~\cite{liu2024physgen} further use LLMs and novel training objectives to enforce physical correctness and semantic consistency from textual prompts.
%, avoiding pixel-space artifacts and producing videos with complex, physically accurate motion. 

\subsection{Physics Enhanced Video Diffusion Models}
Video diffusion models~\cite{ho2020denoising, ho2022video, hong2022cogvideo, opensora} are inherently probabilistic, and while they achieve high visual realism, adherence to physical rules is not guaranteed~\cite{motamed2025generative}. To improve physical consistency, several recent works~\cite{cao2024teaching, lin2025reasoning, lin2025reasoning, chen2024diffphys, lv2024gpt4motion, liu2024physgen} leverage large language models (LLMs) as multi-modal supervisory signals. Other works directly incorporate some physical priors: MotionCraft~\cite{montanaro2024motioncraft} warps the latent noise space of image diffusion models using optical flow from simulations while PhysDiff~\cite{yuan2023physdiff} iteratively projects motion onto physically plausible spaces during generation. These approaches rely on refined prompts, optical flow, or pixel-level projections as implicit physical signals to guide video generation. Although they can improve general physical plausibility, they are often insufficient for enforcing specific physical behaviors in certain scenarios. In contrast, our method explicitly embeds physical states into the video generation model, jointly producing videos and their corresponding physical quantities, which is a capability that existing video generation models cannot achieve.
\vspace{-1.0em} % 可选，调整上下间距

%from low-resolution inputs while maintaining physical accuracy, even for sparse or incomplete observations

\subsection{Deep learning Methods for Fluid Dynamics}
Machine learning has been increasingly applied to accelerate or/and improve the numerical solutions for complex fluid dynamics problems.~\cite{bastek2024physics} combines denoising diffusion models with PINNs, incorporating PDEs constraints during training to reduce residuals. Physics-Informed Neural Network Super-Resolution~\cite{wang2020physics} integrates traditional super-resolution techniques with physics consistency losses, preserving physical accuracy of high-grid resolution data. PirateNets~\cite{wang2024piratenets} introduce physics-informed initialization to improve PINNs trainability and scalability, addressing derivative initialization issues. These approaches demonstrate that using deep learning to solve physical equations or integrating neural networks into traditional solvers can improve the accuracy, and scalability of PDEs solutions, or accelerate the generation of high-grid resolution physics states. However, these methods do not address the bottleneck of the time-intensive rendering required in the visualization stage. 

In contrast, we are the first to address this problem by jointly integrating physics state prediction and rendering into one video diffusion model with faster inference time compared to classical rendering pipelines.

% our approach significantly reduces the time from physical simulation to video output compared to traditional simulation-plus-rendering pipelines, while still maintaining relatively high physical fidelity.

% In contrast, our approach jointly generates the rendered video alongside the physical states. 

\vspace{-1em}
\begin{figure*}[ht]
  \centering
  \vspace{1em}
  \includegraphics[width=0.9\textwidth]{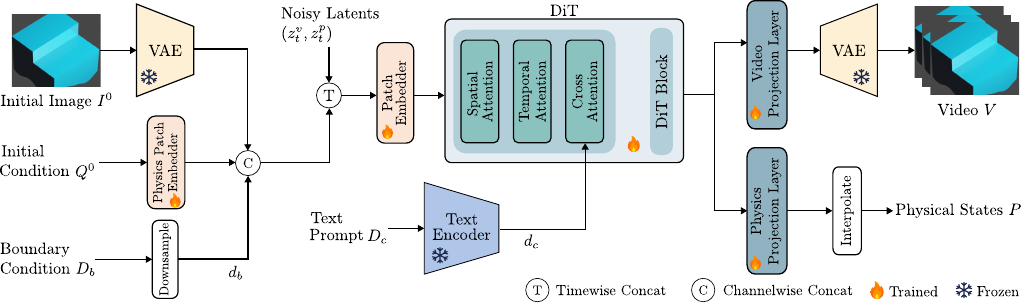}
  \caption{Model architecture. We condition the DiT on the initial and boundary conditions ($Q^0$ and $D_b$) of the SWEs, and a rendered frame $I^0$, to jointly generate the video $V$ and physical states $P$.}
  \label{fig:method}
  \vspace{-1.5em}
\end{figure*}

% Independent noise is added to each modality, then concatenated along the channel dimension. The fused representation is denoised via DiT blocks, and the final outputs are produced by two convolution-based projection layers.
\section{Methods}
\label{sec:pagestyle}

In this section, we first introduce the task formulation, some SWEs preliminaries and the key components of our method. 

%As shown in Fig.~\ref{fig:method}, our framework supports two output modalities, rendered video and corresponding physical states, and incorporates three forms of control signals: initial and boundary conditions of the equations with a text prompt. We first review our Task Formulation (Sec.~\ref{sec:Task Formulation}), followed by the background of the SWEs, which includes the equation itself and the classical Finite Volume Method (FVM) (Sec. ~\ref{sec:background}). Finally, we present our physics-informed video diffusion model (Sec. ~\ref{sec:model}).
\vspace{-1.3em} % 可选，调整上下间距

\subsection{Task Formulation}
\label{sec:Task Formulation}
Our goal is to jointly generate a realistic $N$ frame video 
$V = \{I^f\}_{f=0}^{N-1}$ 
and corresponding physical states $P = \{\hat{Q}^f\}_{f=0}^{N-1}$, where $I, \hat{Q} \in \mathcal{R}^{3\times H\times W}$ given the initial image and physics conditions $I^{0}, Q^{0}$, boundary conditions $D_b \in \mathcal{R}^{H\times W}$ and a text prompt $D_c$.

% \george{should be   }

% \george{notation, and explain what they are, hb height of the water bed, and the dimensions of each one. This should be done for the boundary AND the initial conditions and a rendered initial frame  with text prompt $c$}

\subsection{Background}
\label{sec:background}
% \george{Add units }
\noindent\textbf{Shallow Water Equations (SWEs).} \\
The SWEs are derived from the depth-averaged Navier–Stokes equations~\cite{navierstokes}, neglecting viscosity, turbulence, Coriolis and surface shear terms. They form a system of nonlinear hyperbolic PDEs~\cite{lax2006hyperbolic}. In two dimensions, it can be written
\begin{equation}
\frac{\partial \vec{Q}}{\partial t}+\frac{\partial \vec{F}}{\partial x}+\frac{\partial \vec{G}}{\partial y}=S,
\label{eq:analytical solution}
\end{equation}
% where $h$ is the water depth and $hu, hv$ are the momenta in the $x$ and $y$ directions, respectively.  
% The fluxes in the $x$ and $y$ directions are denoted by $\vec{F}$ and $\vec{G}$, while $\vec{S}$ represents the bed slope source term. 
% The river bottom is described by the profile $S(x,y)$, and $g$ denotes the gravitational constant. Expanding the system gives
where $\vec{Q}=(h,hu,hv)^T$ denotes water height and momentum in $x$ and $y$. The fluxes in the $x$ and $y$ directions are denoted by $\vec{F}$ and $\vec{G}$, while $S$ is the bed slope source term, which also represents the boundary condition. The river bottom is specified by a profile $S(x, y)$, where $g$ is the gravitational constant. Expanding yields

% \george{Here introduce what Db is- the boundary condition}.
\begin{equation}
\begin{aligned}
\frac{\partial}{\partial t}\left(\begin{array}{c}
h \\
h u \\
h v
\end{array}\right)+\frac{\partial}{\partial x}\left(\begin{array}{c}
h u \\
h u^2 +\frac{1}{2} g h^2 \\
h u v
\end{array}\right)\\
+\frac{\partial}{\partial y}\left(\begin{array}{c}
h v\\
h u v \\
h v^2 +\frac{1}{2} g h^2
\end{array}\right)
=\left(\begin{array}{c}
0 \\
-g h \frac{\partial S}{\partial x}(x, y) \\
-g h \frac{\partial S}{\partial y}(x, y)
\end{array}\right)
\end{aligned}
\end{equation}
Solutions of the SWEs may exhibit discontinuities, and evolution at cell interfaces can be interpreted as a Riemann problem~\cite{toro2013riemann}, showing how a discontinuity propagates over time.

\noindent\textbf{Finite Volume Method (FVM) for Hyperbolic PDEs.}\\ The FVM~\cite{eymard2000finite, leveque2002finite, toro2001shock} evolves cell averages over control volumes, ensuring discrete conservation and properly handling discontinuities. On a uniform mesh, the cell average at spatial location $i$ and time step $n$ is
\begin{equation}
Q_i^n \approx \frac{1}{\Delta x}\int_{\Omega_i} Q(x,t^n)\,dx,
\end{equation}
integrating over $\Omega_i \times [t^n,t^{n+1})$ gives the numerical solution
\begin{equation}
Q_i^{n+1} = Q_i^n - \frac{\Delta t}{\Delta x}\Big(\bar{F}_{i+1/2}^n - \bar{F}_{i-1/2}^n\Big),
\label{eq:numerical solution}
\end{equation}
where $\bar{F}_{i\pm 1/2}^n$ are the time-averaged numerical fluxes at cell interfaces. These fluxes are computed using Riemann solver methods such as Lax–Friedrichs, Rusanov, or Roe~\cite{toro2013riemann}, which approximate the solution of the local Riemann problem and ensure stable propagation of waves across discontinuities.
\vspace{-1em} % 可选，调整上下间距
\subsection{Physics-informed video diffusion model}
\vspace{-0.5em}

\label{sec:model} 

\textbf{Model Overview.} Our model is an image conditioned multi-modal Latent Diffusion Model (LDM) that generates two output modalities: the rendered video and corresponding physical states, given three input conditions: initial image and physics conditions, the boundary conditions and a text prompt. The physical states are chosen to be equal to the full time-series of FVM solutions $Q_i^n$ of Eq.~\ref{eq:numerical solution} at each point. 

% , where $h$ is water height and $u,v$ are velocity components).

% After discretizing the SWEs (Eq.~\ref{eq:numerical solution}), the full time-series of FVM solutions $Q_i^n$ at each grid point of Eq.~\ref{eq:analytical solution} is obtained via ODE integration. 

 \noindent\textbf{Diffusion Model Training.} 
 A pre-trained Variational Autoencoder (VAE) maps videos to latents $z_{v} \in \mathcal{R}^{4 \times N \times H' \times W'}$, where $H'$ and $W'$ are the video dimensions after the VAE downsampling. We process the physics states with a patch embedding layer to the same spatial resolution of the video latents, obtaining $z_{p} \in \mathcal{R}^{3 \times N \times H' \times W'}$. The boundary conditions $D_b$ are interpolated to the latent space dimensions obtaining $d_b \in \mathcal{R}^{N \times H' \times W'}$. The text prompt is encoded with the T5 text encoder~\cite{raffel2020exploring} to obtain a caption latent $d_c$. The set of input conditions is denoted by $\mathcal{C} = \{z_p^0, z_v^0, d_b,  d_c\}$.
 
 We then apply a $T$-step Gaussian noising process to both $z^v$ and $z^p$, producing noisy video and noisy physical states  $z^v_t$ and $z^p_t$ that are progressively denoised by a Diffusion Transformer (DiT) network $\epsilon_\theta(\cdot, t)$. Training is performed with a joint objective $\mathcal{L}_{\text{total}} = \mathcal{L}_{\text{video}} + \mathcal{L}_{\text{phys}}$, where
\[
\mathcal{L}_{\text{video}} = 
\mathcal{E}_{z^{v},\mathcal{D},\epsilon\sim\mathcal{N}(0,1),t}
\left\|\epsilon - \epsilon_\theta(z^v_t, \mathcal{C}, t)\right\|_2^2
\]
\[
\mathcal{L}_{\text{phys}} = 
\mathcal{E}_{z^{p},\mathcal{D},\epsilon\sim\mathcal{N}(0,1),t}
\left\|\epsilon - \epsilon_\theta(z^{p}_t, \mathcal{C}, t)\right\|_2^2
\]
\textbf{Model Architecture.} As shown in Fig.~\ref{fig:method}, a physics embedding layer encodes the initial conditions $Q^0$ to the same spatial resolution as the video latent $z^v$. Forward diffusion is applied independently to the video and physics latents ($z^v$ and $z^p$), i.e. independent noise is added to each modality. These noisy latents are then concatenated along the channel dimension with the boundary conditions $d_b$, ensuring that physical constraints are incorporated into the fused representation. The combined latent along with the injected prompt latents $d_c$, are passed through a series of DiT blocks, which perform spatio-temporal denoising while leveraging the physics-informed features. Finally, two separate Convolutional neural network (CNN)-based projection heads ($P_v$ and $P_p$) map the denoised representation to the video and physics latents, respectively, allowing the model to jointly generate visually realistic video frames and physically consistent states.

\vspace{-1.5em}
\section{Experiments}
\vspace{-1em} % 可选，调整上下间距

\label{sec:typestyle}
To evaluate our physics-informed video diffusion model for the SWEs, we focus on its two main contributions: higher quality than normal video diffusion models and faster generation than traditional simulation-plus-rendering pipelines. \\
\noindent\textbf{Datasets.} 
Training and evaluation data were generated with a classical 2D Riemann solver using a second-order Roe flux from Clawpack~\cite{mandli2016clawpack} with periodic boundary conditions. Initial conditions were randomly sampled to produce 20K simulations with diverse waterbeds and 10K with a planar riverbed, each on $128\times128$, $256\times256$, and $512\times512$ grids over 1.5 seconds using TVD Runge–Kutta method~\cite{gottlieb1998total}. And all the videos were rendered by Blender~\cite{blain2019complete}.\\
\noindent\textbf{Baseline Models.} 
We compared our method with several video generation models: CogVideoX-Fun, CogVideoX (I2V)-LoRA~\cite{hong2022cogvideo}, and OpenSora-1.1~\cite{opensora} (also can be one ablation), all fine-tuned in our dataset under comparable settings. As no previous work has addressed this combination of objectives, our experiments aim to demonstrate superior video quality compared to state-of-the-art diffusion models and faster generation than classical solvers. Therefore, other PINN-based solvers are not suitable baselines.\\
\begin{figure}[!t]
  \centering
  \includegraphics[width=0.45\textwidth]{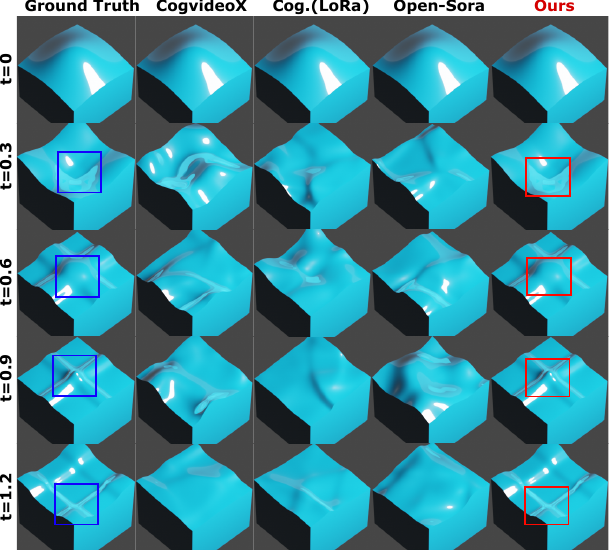} % 调整宽度到半栏
  \caption{Qualitative comparison of Gaussian bumps results.}
  \vspace{-1em}
  \label{fig:main_comparison}
  \vspace{-1em} % 可选，调整上下间距
\end{figure}
\noindent\textbf{Ablation Study.} To thoroughly evaluate our framework, we perform ablation studies with four variants: (i) naive video diffusion model without physics-informed input, and three physics-informed models using different physics patch embedding strategies for encoding physical states: (ii) linear interpolation (LI.), (iii) CNN-based embeddings, and (iv) multilayer perceptron (MLP)-based embeddings. This setup allows us to systematically investigate the impact of physics conditioning and the choice of embedding on video quality.

\noindent\textbf{Implementation Details.} Our physics-informed, image-conditioned video diffusion model is based on the pre-trained OpenSora-1.1~\cite{opensora}. Training and inference are performed at resolutions matching the datasets, with 21 frames. Inference uses 50 sampling steps. The model is optimized using AdamW~\cite{loshchilov2017decoupled} for 20K steps, taking about one day on a single NVIDIA A800 GPU.

\noindent\textbf{Evaluation.}
We evaluate on 50 test cases: 30 with random initial conditions over varying terrains, 10 planar riverbeds, 5 Gaussian bumps, and 5 classical dam-breaks. Evaluation metrics are:

% ...

\begin{itemize}[itemsep=0em, topsep=0em, parsep=0em, partopsep=0em, leftmargin=*]
\item Video Quality: we evaluate video fidelity using LPIPS, FVD, PSNR, and SSIM against ground-truth renderings.
\item Physics Accuracy: we compute the mean $L_1$ norm loss of $(h, hu, hv)$ over all time steps against the output of the classical solver.

\item Time Consumption: we compare total computation time between our method and the classical pipeline, where classical time includes simulation (Sim.) and rendering.
\end{itemize}

% \george{L2 Loss ? }

\begin{table}[!t]
    \centering
    \setlength{\tabcolsep}{3pt}
    \renewcommand\arraystretch{1}
    \begin{adjustbox}{width=0.95\columnwidth}
    \begin{tabular}{l|cccc}
    \hline
    Method & LPIPS$\downarrow$ & SSIM$\uparrow$ & PSNR$\uparrow$ & FVD$\downarrow$ \\
    \hline
    CogVideoX-Fun & 0.2262 & 0.7994 & 18.63 & 189.53 \\
    CogVideoX (I2V)-LoRA & 0.2241 & 0.8036 & 18.89 & 178.37 \\
    Naive without Physics & 0.2411 & 0.7862 & 18.28 & 192.64 \\
    \hline
    LI. with Physics & 0.1588 & 0.8355 & 22.19 & 137.20 \\
    MLP with Physics & 0.1366 & 0.8423 & 24.91 & 128.69 \\    
    \textbf{CNN with Physics} & \textbf{0.1341} & \textbf{0.8519} & \textbf{25.86} & \textbf{125.13} \\
    \hline
    \end{tabular}
    \end{adjustbox}
    \vspace{-0.5em}

    \caption{Main comparison metrics of video quality across video generation methods. Bold indicates best results. Upper: Baselines, Lower: Ablations}
    \label{tab:main_comparison}
    \vspace{-0.5em}

\end{table}

% \george{Can you compare with any method cited in the introduction ? if no method can do the same as you do, then you should say you are the first to do this. Also you should compare this with a classical approach. You will be worse, but you can mention how much time you saved compared to the loss of accuracy. One thing you can do is a grah of perfromance versus time, to show your contribution.}
\begin{table}[!t]
\centering
\vspace{-0.5em}
\small

\resizebox{0.45\textwidth}{!}{%
\begin{tabular}{c|c|ccc|c}  
\hline
\multirow{2}{*}{Resolution} & \multirow{2}{*}{Method} & \multicolumn{3}{c|}{Time (s)} & \multirow{2}{*}{\makecell{Accuracy\\(\%)}} \\
 & & Sim. & Render & Total & \\
\hline
128$\times$128 & Classical & 5.6 & 572 & 577.6 & - \\
128$\times$128 & \textbf{Ours}       & -    & -    & \textbf{12} & 90.2 \\
256$\times$256 & Classical & 10.3 & 788 & 798.3 & - \\
256$\times$256 & \textbf{Ours}       & -    & -    & \textbf{15} & 73.6 \\
512$\times$512 & Classical & 18.9 & 1463 & 1481.9 & - \\
512$\times$512 & \textbf{Ours}       & -    & -    & \textbf{18} & 67.1 \\
\hline
\end{tabular}%
}
\caption{Comparison of classical (Classical) pipleline, Clawpack with Blender  and our CNN with Physics method in different grid resolutions.}
\label{tab:comparison}
\vspace{-1.5em}
\end{table}

% \input{Tables/image_editing}

% \input{Figures/Image_Editing_Comparison}
% \label{sec:majhead}
\vspace{-0.4em}
\noindent\textbf{Discussion} From Fig.~\ref{fig:main_comparison}, we observe that models without physics inputs produce nearly random wave variations, whereas our method closely matches the ground truth and accurately captures wave dynamics. This observation is further supported by Tab.~\ref{tab:main_comparison}, which shows that our method outperforms the baselines on visual metrics. The ablation results in the same table indicate that the CNN-based embedder generally outperforms both LI. and MLP-based embedders. On the other hand, Tab.~\ref{tab:comparison} highlights efficiency: traditional simulation-plus-rendering pipelines see a significant increase in computation and rendering time as grid resolution rises, while our inference time remains nearly constant. One limitation of our model is the decreased accuracy of physical states as the resolution increases. Our approach can be combined with stronger video foundation models to increase the accuracy at high resolutions.

\vspace{-1.5em} % 可选，调整上下间距
\section{Conclusion}
\vspace{-1em} % 可选，调整上下间距

\label{sec:print}
We proposed a physics-informed video generation framework that incorporates grid-based method into diffusion models, enabling the co-generation of videos and physical states under the SWEs. Our approach bypasses expensive rendering, yet achieves visual quality comparable to classical simulation-plus-rendering pipelines while producing results orders of magnitude faster. Compared with purely data-driven video diffusion, our method also enforces physical consistency, yielding temporally stable and interpretable outputs.

Despite these advantages, limitations remain. First, the accuracy of generated physical states degrades at higher resolutions. Second, our current work is restricted to the SWEs; extending the framework to more general governing equations, such as the Euler equations, represents an important direction for future research.

\vfill\pagebreak

% \section{REFERENCES}
% \label{sec:refs}

% List and number all bibliographical references at the end of the
% paper. The references can be numbered in alphabetic order or in
% order of appearance in the document. When referring to them in
% the text, type the corresponding reference number in square
% brackets as shown at the end of this sentence \cite{C2}. An
% additional final page (the fifth page, in most cases) is
% allowed, but must contain only references to the prior
% literature.

% Please follow the IEEE Citation Guidelines, \url{https://ieee-dataport.org/sites/default/files/analysis/27/IEEE\%20Citation\%20Guidelines.pdf} for formatting of references.

% \bibliographystyle{IEEEbib}
% \bibliography{strings,refs}

\begingroup
\setstretch{1.0}
\setlength\bibitemsep{0pt}
\printbibliography

@article{navierstokes,
  title={Navier--stokes revisited},
  author={Brenner, Howard},
  journal={Physica A: Statistical Mechanics and its Applications},
  volume={349},
  number={1-2},
  pages={60--132},
  year={2005},
  publisher={Elsevier}
}

@article{johanson2004real,
  title={Real-time water rendering},
  author={Johanson, Claes and Lejdfors, Calle},
  journal={Lund University},
  year={2004}
}

@article{ho2020denoising,
  title={Denoising diffusion probabilistic models},
  author={Ho, Jonathan and Jain, Ajay and Abbeel, Pieter},
  journal={Advances in neural information processing systems},
  volume={33},
  pages={6840--6851},
  year={2020}
}

@article{ho2022video,
  title={Video diffusion models},
  author={Ho, Jonathan and Salimans, Tim and Gritsenko, Alexey and Chan, William and Norouzi, Mohammad and Fleet, David J},
  journal={Advances in neural information processing systems},
  volume={35},
  pages={8633--8646},
  year={2022}
}

@inproceedings{liu2024physgen,
  title={Physgen: Rigid-body physics-grounded image-to-video generation},
  author={Liu, Shaowei and Ren, Zhongzheng and Gupta, Saurabh and Wang, Shenlong},
  booktitle={European Conference on Computer Vision},
  pages={360--378},
  year={2024},
  organization={Springer}
}

@article{montanaro2024motioncraft,
  title={Motioncraft: Physics-based zero-shot video generation},
  author={Montanaro, Antonio and Savant Aira, Luca and Aiello, Emanuele and Valsesia, Diego and Magli, Enrico},
  journal={Advances in Neural Information Processing Systems},
  volume={37},
  pages={123155--123181},
  year={2024}
}

@article{thurey2010multiscale,
  title={A multiscale approach to mesh-based surface tension flows},
  author={Th{\"u}rey, Nils and Wojtan, Chris and Gross, Markus and Turk, Greg},
  journal={ACM Transactions on Graphics (TOG)},
  volume={29},
  number={4},
  pages={1--10},
  year={2010},
  publisher={ACM New York, NY, USA}
}

@article{wojtan2010physics,
  title={Physics-inspired topology changes for thin fluid features},
  author={Wojtan, Chris and Th{\"u}rey, Nils and Gross, Markus and Turk, Greg},
  journal={ACM Transactions on Graphics (TOG)},
  volume={29},
  number={4},
  pages={1--8},
  year={2010},
  publisher={ACM New York, NY, USA}
}

@article{foster1996realistic,
  title={Realistic animation of liquids},
  author={Foster, Nick and Metaxas, Dimitri},
  journal={Graphical models and image processing},
  volume={58},
  number={5},
  pages={471--483},
  year={1996},
  publisher={Elsevier}
}

@inproceedings{stam2023stable,
  title={Stable fluids},
  author={Stam, Jos},
  booktitle={Seminal Graphics Papers: Pushing the Boundaries, Volume 2},
  pages={779--786},
  year={2023}
}

@inproceedings{premvzoe2003particle,
  title={Particle-based simulation of fluids},
  author={Prem{\v{z}}oe, Simon and Tasdizen, Tolga and Bigler, James and Lefohn, Aaron and Whitaker, Ross T},
  booktitle={Computer Graphics Forum},
  volume={22},
  pages={401--410},
  year={2003},
  organization={Wiley Online Library}
}

@inproceedings{solenthaler2009predictive,
  title={Predictive-corrective incompressible SPH},
  author={Solenthaler, Barbara and Pajarola, Renato},
  booktitle={ACM SIGGRApH 2009 papers},
  pages={1--6},
  year={2009}
}

@article{eymard2000finite,
  title={Finite volume methods},
  author={Eymard, Robert and Gallou{\"e}t, Thierry and Herbin, Rapha{\`e}le},
  journal={Handbook of numerical analysis},
  volume={7},
  pages={713--1018},
  year={2000},
  publisher={Elsevier}
}

@book{toro2013riemann,
  title={Riemann solvers and numerical methods for fluid dynamics: a practical introduction},
  author={Toro, Eleuterio F},
  year={2013},
  publisher={Springer Science \& Business Media}
}

@book{leveque2002finite,
  title={Finite volume methods for hyperbolic problems},
  author={LeVeque, Randall J},
  volume={31},
  year={2002},
  publisher={Cambridge university press}
}

@book{toro2001shock,
  title={Shock-capturing methods for free-surface shallow flows},
  author={Toro, Eleuterio Francisco and others},
  year={2001},
  publisher={Wiley and Sons Ltd.}
}

@article{gottlieb1998total,
  title={Total variation diminishing Runge-Kutta schemes},
  author={Gottlieb, Sigal and Shu, Chi-Wang},
  journal={Mathematics of computation},
  volume={67},
  number={221},
  pages={73--85},
  year={1998}
}

@article{opensora,
  title={Open-sora: Democratizing efficient video production for all},
  author={Zheng, Zangwei and Peng, Xiangyu and Yang, Tianji and Shen, Chenhui and Li, Shenggui and Liu, Hongxin and Zhou, Yukun and Li, Tianyi and You, Yang},
  journal={arXiv preprint arXiv:2412.20404},
  year={2024}
}

@article{loshchilov2017decoupled,
  title={Decoupled weight decay regularization},
  author={Loshchilov, Ilya and Hutter, Frank},
  journal={arXiv preprint arXiv:1711.05101},
  year={2017}
}

@article{motamed2025generative,
  title={Do generative video models understand physical principles?},
  author={Motamed, Saman and Culp, Laura and Swersky, Kevin and Others},
  journal={arXiv preprint arXiv:2501.09038},
  year={2025}
}

@article{cao2024teaching,
  title={Teaching video diffusion model with latent physical phenomenon knowledge},
  author={Cao, Qinglong and Wang, Ding and Li, Xirui and Chen, Yuntian and Ma, Chao and Yang, Xiaokang},
  journal={arXiv preprint arXiv:2411.11343},
  year={2024}
}

@article{lin2025reasoning,
  title={Reasoning physical video generation with diffusion timestep tokens via reinforcement learning},
  author={Lin, Wang and Jia, Liyu and Hu, Wentao and Pan and Others},
  journal={arXiv preprint arXiv:2504.15932},
  year={2025}
}

@article{chen2024diffphys,
  title={DiffPhys: enhancing signal-to-noise ratio in remote photoplethysmography signal using a diffusion model approach},
  author={Chen, Shutao and Wong, Kwan-Long and Chin, Jing-Wei and Chan, Tsz-Tai and So, Richard HY},
  journal={Bioengineering},
  volume={11},
  number={8},
  pages={743},
  year={2024},
  publisher={MDPI}
}

@inproceedings{lv2024gpt4motion,
  title={Gpt4motion: Scripting physical motions in text-to-video generation via blender-oriented gpt planning},
  author={Lv, Jiaxi and Huang, Yi and Yan, Mingfu and Huang and Others},
  booktitle={Proceedings of the IEEE/CVF conference on computer vision and pattern recognition},
  pages={1430--1440},
  year={2024}
}

@article{kochkov2021machine,
  title={Machine learning--accelerated computational fluid dynamics},
  author={Kochkov, Dmitrii and Smith, Jamie A and Alieva, Ayya and Wang, Qing and Brenner, Michael P and Hoyer, Stephan},
  journal={Proceedings of the National Academy of Sciences},
  volume={118},
  number={21},
  pages={e2101784118},
  year={2021},
  publisher={National Academy of Sciences}
}

@inproceedings{yuan2023physdiff,
  title={Physdiff: Physics-guided human motion diffusion model},
  author={Yuan, Ye and Song, Jiaming and Iqbal and Others},
  booktitle={Proceedings of the IEEE/CVF international conference on computer vision},
  pages={16010--16021},
  year={2023}
}

@article{cai2021physics,
  title={Physics-informed neural networks (PINNs) for fluid mechanics: A review},
  author={Cai, Shengze and Mao, Zhiping and Wang, Zhicheng and Others},
  journal={Acta Mechanica Sinica},
  volume={37},
  number={12},
  pages={1727--1738},
  year={2021},
  publisher={Springer}
}

@article{bastek2024physics,
  title={Physics-informed diffusion models},
  author={Bastek, Jan-Hendrik and Sun, WaiChing and Kochmann, Dennis M},
  journal={arXiv preprint arXiv:2403.14404},
  year={2024}
}

@article{wang2020physics,
  title={Physics-informed neural network super resolution for advection-diffusion models},
  author={Wang, Chulin and Bentivegna, Eloisa and Zhou, Wang and Klein, Levente and Elmegreen, Bruce},
  journal={arXiv preprint arXiv:2011.02519},
  year={2020}
}

@article{wang2024piratenets,
  title={Piratenets: Physics-informed deep learning with residual adaptive networks},
  author={Wang, Sifan and Li, Bowen and Chen, Yuhan and Perdikaris, Paris},
  journal={Journal of Machine Learning Research},
  volume={25},
  number={402},
  pages={1--51},
  year={2024}
}

@book{lax2006hyperbolic,
  title={Hyperbolic partial differential equations},
  author={Lax, Peter D},
  volume={14},
  year={2006},
  publisher={American Mathematical Soc.}
}

@article{hong2022cogvideo,
  title={CogVideo: Large-scale Pretraining for Text-to-Video Generation via Transformers},
  author={Hong, Wenyi and Ding, Ming and Zheng, Wendi and Liu, Xinghan and Tang, Jie},
  journal={arXiv preprint arXiv:2205.15868},
  year={2022}
}

@article{jasak2009openfoam,
  title={OpenFOAM: Open source CFD in research and industry},
  author={Jasak, Hrvoje},
  journal={International journal of naval architecture and ocean engineering},
  volume={1},
  number={2},
  pages={89--94},
  year={2009},
  publisher={Elsevier}
}

@article{mandli2016clawpack,
  title={Clawpack: building an open source ecosystem for solving hyperbolic PDEs},
  author={Mandli, Kyle T and Ahmadia, Aron J and Berger, Marsha and Calhoun, Donna and George, David L and Hadjimichael, Yiannis and Ketcheson, David I and Lemoine, Grady I and LeVeque, Randall J},
  journal={PeerJ Computer Science},
  volume={2},
  pages={e68},
  year={2016},
  publisher={PeerJ Inc.}
}

@article{kenneally2020basilisk,
  title={Basilisk: A flexible, scalable and modular astrodynamics simulation framework},
  author={Kenneally, Patrick W and Piggott, Scott and Schaub, Hanspeter},
  journal={Journal of aerospace information systems},
  volume={17},
  number={9},
  pages={496--507},
  year={2020},
  publisher={American Institute of Aeronautics and Astronautics}
}

@article{raffel2020exploring,
  title={Exploring the limits of transfer learning with a unified text-to-text transformer},
  author={Raffel, Colin and Shazeer, Noam and Roberts, Adam and Lee, Katherine and Narang, Sharan and Matena, Michael and Zhou, Yanqi and Li, Wei and Liu, Peter J},
  journal={Journal of machine learning research},
  volume={21},
  number={140},
  pages={1--67},
  year={2020}
}

@book{blain2019complete,
  title={The complete guide to Blender graphics: computer modeling \& animation},
  author={Blain, John M},
  year={2019},
  publisher={AK Peters/CRC Press}
}
\endgroup

% % \bibliographystyle{IEEEbib}
% \bibliographystyle{IEEEtran}

% \bibliography{strings,refs}

\end{document}